\documentclass[aps,prd,reprint,superscriptaddress,nofootinbib]{revtex4-1}

\usepackage{comment}
\usepackage[utf8]{inputenc} 
\usepackage{graphicx,color,overpic,mathtools}
\usepackage{amsthm,amsmath,amssymb,hyperref,mathrsfs}
\usepackage{braket,bm,bbm,setspace}
\PassOptionsToPackage{normalem}{ulem}
\usepackage{ulem} 
\usepackage{physics}
\usepackage{float}
\usepackage[makeroom]{cancel}
\usepackage[english]{babel}
\usepackage{graphicx}
\graphicspath{ {images/} }
\addto\captionsspanish{}
\hypersetup{
    colorlinks=true,
    pdfborder={0 0 0},
    linkcolor=blue,
    citecolor=blue,
    urlcolor=blue
}
\definecolor{rRGB}{RGB}{255, 0, 0}
\begin{document}
\title{Can light-rings self-gravitate?}
\author{Francesco Di Filippo}
\affiliation{Institute of Theoretical Physics, Faculty of Mathematics and Physics, Charles University, V Holešovičkách 2, 180 00 Prague 8, Czech Republic}
\author{Luciano Rezzolla }
\affiliation{Institut f\"ur Theoretische Physik, Max-von-Laue-Strasse 1, 60438 Frankfurt, Germany}
\affiliation{Frankfurt Institute for Advanced Studies, Ruth-Moufang-Strasse 1, 60438 Frankfurt, Germany}
\affiliation{School of Mathematics, Trinity College, Dublin 2, Ireland}

\begin{abstract}
In a spherically symmetric and static spacetime of a compact object, such as that of a Schwarzschild black hole, the light-ring is a 2-sphere where photons experience the only possible circular orbits. As a ``Gedankenexperiment", we imagine an advanced civilisation able to populate the light-ring of a nonrotating black hole of mass $M$ with photons having a fine-tuned impact parameter that allows their orbits to be exactly circular with radius $r=3M$. As the number of photons in the light-ring increases in time, its mass will no longer be negligible and hence it will impact on the background spacetime, that is, it will ``self-gravitate". We here consider two different routes to assign a nonzero mass to the light-ring that are either based on a discrete concentration of photons on a specific radial location or on a suitable distribution of photons in a given region. In both cases, and using the Einstein equations, we find that the inclusion of the energy from the accumulated photons leads to the generation of new light-rings. Such new light-rings can either appear at well-defined but discrete locations, or be fused in a well-defined region. In either case, we show that such light-ring configurations are dynamically unstable and a small perturbation, either via the inclusion of an additional photon onto the light-ring or via the absorption of a photon by the black hole, leads to a catastrophic destruction of the light-ring structures.
\end{abstract}
\maketitle

%------------------------------------------------------
%\section{Introduction}
%\label{sec:intro}
%------------------------------------------------------

\smallskip
\noindent\textit{Introduction.~} Light-rings, namely, the location of unstable circular orbits, play a crucial role in the study of astrophysical compact objects as shown by a number of a theoretical investigations exploring the spacetime on those scales (see, e.g.,~\cite{Chirenti:2007mk, Vincent:2015xta, Glampedakis:2017dvb, Cardoso:2017cfl, Jai-akson:2017ldo, Glampedakis:2017cgd, Cardoso:2016rao, Carballo-Rubio:2018jzw, Cardoso:2019rvt, Volkel:2020xlc, Kocherlakota2021, Volkel:2022khh, Eichhorn:2022oma, Moreira:2024sjq, Pedrotti:2024znu, Genzel:2024vou}), but also by the observations of supermassive black holes~\cite{EventHorizonTelescope:2019dse, EHT_Sgr, EventHorizonTelescope:2022xqj, Kocherlakota2022} or through the ringdown associated with the  gravitational-wave signal from merging black-hole binaries~\cite{PhysRevLett.116.061102, LIGOScientific:2019fpa, LIGOScientific:2020tif, LIGOScientific:2021sio}. Indeed, it is generally accepted that the study of light-rings, either via black-hole imaging or via black-hole quasi-normal modes, represents one of the most promising routes to test theories of gravity beyond general relativity. 

Interestingly, while light-rings in general relativity are well understood, both for nonrotating and rotating black holes \cite{Khoo:2016xqv,Cunha:2020azh,Hod:2022mys}, but also for other compact objects not possessing an event horizon \cite{Cunha2017,Franzin:2023slm,DiFilippo:2024mnc}, and have been the subject of uncountable works, including works aimed at the study of populated spacetimes \cite{Andreasson:2015agw,Andreasson:2021lsh,Macedo:2024qky}, there is an question that so far has escaped the attention of the literature, namely, \textit{``can light-rings self-gravitate?"} 

This work is dedicated to addressing this question by carrying out a ``Gedankenexperiment'' to explore the behaviour of the light-ring of a Schwarzschild black hole. More specifically, we postulate the existence of an advanced civilisation that is able to approach a (nonrotating) black hole and has developed a technology that allows it to shoot large numbers of photons towards the black hole with the exact impact parameter corresponding to the unstable circular orbit placed at $r=3\,M$, where $M$ is the mass of the black hole. In this way, the civilisation is able to carefully populate the light-ring with an arbitrarily large number of photons, all having the necessary fine-tuned impact parameter $b=\sqrt{27}M$. 

However, as the number of photons in the light-ring increases with time, and ignoring any photon-photon interaction, its mass will no longer be negligible and hence it will start to ``self-gravitate", that is, it will have a sizable impact on the background spacetime\footnote{Photon-photon will tend to destabilise photons previously present on the light-ring, but as long as the lost photons are replaced by new and more numerous photons, the mass of the light-ring will grow.}. Under these conditions, the dynamics of the photons can no longer be considered as that in the original black-hole spacetime in vacuum, but it will be regulated by the geometry of the spacetime resulting from the proper solution of the Einstein equations for a black hole surrounded by the energy contributions coming from the thin null shells of photons on (unstable) circular orbits. 

To address this scenario, we consider two different routes. In the first one, which we refer to as the ``discrete sequence", we show that the solution of the Einstein equations leads to the formation of a new light-ring located outside the first one and at a position that depends on the mass of the light-ring. In full analogy with the first light-ring, also the second light-ring can be populated with photons and acquire a mass, thus leading to the appearance of a third light-ring placed outside the second light-ring at a radial location that will again depend on the total energy of the photons. Clearly, the advanced civilisation can iterate this procedure for an arbitrarily number of times  and hence construct a whole hierarchy of discrete, nested, and self-gravitating light-rings that are increasingly large in size. However, adding a single photon on any of the light-rings would change the geometry of the constructed spacetime and hence destabilise all the photons on the various discrete light-rings, leading to a catastrophic destruction of the hierarchy. In the second scenario, we do not consider a series of discrete shells of light-rings but a suitably constructed continuum distribution of energy in which the shells are infinitesimally separated and thus lead to a region of densely packed light-rings, namely, a ``light-region". Also in this case, however, the fine-tuned series of light rings is unstable once any additional photon is added.

\smallskip
\noindent\textit{Discrete sequence of light-rings.~} We start by recalling that for an axisymmetric and stationary spacetime described by spherical polar coordinates $(t,r,\theta,\phi)$, $\partial_t$ and $\partial_\phi$ are Killing vectors. Light-rings in such axisymmetric and stationary spacetimes containing a black hole are then defined as regions of the spacetime in which null geodesics only have momenta along such Killing vectors, namely, along the $t$ and $\phi$ directions. If we further restrict to a spherically symmetric configuration, which is the interest of this paper, we can write the metric as
\begin{equation}    
ds^2=g_{tt}dt^2+g_{rr}dr^2+r^2d\Omega^2 \,,
\end{equation}
and the locations of the light-rings can be obtained by studying the Hamiltonian for a massless particle with four-momentum $p_{\mu}$, $\mathcal{H}:=g^{\mu\nu} p_\mu p_\nu$.
%
\begin{comment}
    
\begin{equation}\label{eq:Hamiltonian}
    \mathcal{H}:=g^{\mu\nu} p_\mu p_\nu=g^{rr}p_r^2+g^{\theta\theta} p_\theta^2+V(r,\theta)=0\,.
\end{equation}
\end{comment}
%
More specifically, given the potential $V(r,\theta)$
\begin{equation}\label{eq:def_V}
    V(r,\theta):=\frac{E^2}{g_{tt}}+\frac{L^2}{r^2\sin^2\theta} \,,
\end{equation}
where $E:=-p_t$ and $L:=p_\phi$ are constants of motion, the Hamilton equations read
\begin{equation}
    \dot{p}_\mu=-\partial_\mu\mathcal{H}=-\left(\partial_\mu\left(g^{rr}p_r^2\right)+\partial_\mu\left(g^{\theta\theta}p_\theta^2\right)+\partial_\mu V(r,\theta)\right)\,,
\end{equation}
where the dot refers to a derivative with respect to the affine parameter, imply that the conditions for the occurrence of light-rings, i.e., $p_r=p_\theta=\dot{p}_\mu=0$, are equivalent to $V(r,\theta)= 0$ and $\partial_\mu V(r,\theta)= 0$. Furthermore, motion on the light-ring will be stable if its location corresponds to a minimum of the potential and unstable if it corresponds to a maximum or a saddle point.

Rather than considering the potential, it is easier to determine the position of the light-rings by restricting to the equatorial plane and studying the stationary points of the effective potential $H(r):=-{g_{tt}}/{g_{\phi\phi}}$ that does not depend on $E$ or $L$ (see e.g., \cite{Cunha:2016bjh, Cunha:2017qtt, DiFilippo:2024mnc}). This approach reveals, for example, that a single light-ring exists in the case of a Schwarzschild black hole and it is located at $r=3M$, thus marking the position of the only circular but unstable photon orbit. 

Our next step consists of adding a discrete set of shells on a Schwarzschild background. In particular, in our ``Gedankenexperiment" we imagine that the advanced civilisation mentioned above is able to populate the light-ring of a Schwarzschild black hole of initial mass $M_0$ with photons having a fine-tuned impact parameter that allows their orbits to be circular at the light-ring.  We assume that the angular components of the momentum of each photon are uniformly distributed, so that the corresponding light-ring will be a 2-sphere, thus preserving the spherical symmetry of the spacetime. As the number of photons in the light-ring increases, the energy of the shell of photons will no longer be negligible, impacting the background spacetime and hence leading to a new spacetime metric that is no longer that of the Schwarzschild black hole with mass $M_0$. In order to obtain the resulting metric, we assume that the new metric is actually composed of a black hole and a discrete series of $N$ nested null shells of photons on (unstable) circular orbits. This allows us to consider a fully generic approach but it is trivial to restrict the relevant results to the case of a single shell, i.e., for $N=1$. 

More specifically, because the spacetime in between the shells is in vacuum, we can invoke Birkhoff's theorem~\cite{Birkoff} to deduce that the solution of the Einstein equations in this case yields a series of $N+1$ spacetime regions and for each of them the line element, $ds^2_n$, is given by  (see also~\cite{Berry:2022zwv}) 
\begin{equation}
    \label{eq:metric_n_shells}
    ds^2_n = -\xi^2_n\left( 1 - \frac{2M_n}{r} \right)dt^2+\left( 1 - \frac{2M_n}{r} \right)^{-1}dr^2+r^2d\Omega^2\,,
\end{equation}
where $n\in\{0,\dots,N\}$ and $\xi^2_n$ is a set of coefficients needed to guarantee the continuity between the different portions of the spacetime and where $\xi^2_0 =1$. Clearly, the region $n=0$ corresponds to the section between the black hole and the first shell, while the $n$-th region corresponds to the section between the $n$-th and the $(n+1)$-th shell. Obviously, the region outside the $n$-th shell is described by a Schwarzschild spacetime with mass $M_n$ (note that because of our choice of time coordinate, the Killing vector $\boldsymbol{\partial_t}$ does not have unit norm at spatial infinity). 

The mass parameter $M_n$ in Eq.~\eqref{eq:metric_n_shells} represents the total mass enclosed within the corresponding region and is thus given by $M_n := M_0 +\sum_{i=1}^n\delta m_i= M_0 + \delta M_n$,
%
%\begin{equation}\label{eq:M}
%    M_n := M_0 +\sum_{i=1}^n\delta m_i= M_0 + \delta M_n\,,
%\end{equation}
%
where $\delta m_i$ indicates the mass of a single shell and $\delta M_n$ denotes the sum of the masses of the first $n$ shells, i.e., $\delta M_n:=\sum_{i=1}^n\delta m_i$.
%
%\begin{equation}
%    \delta M_n:=\sum_{i=1}^n\delta m_i\,.
%\end{equation}

\begin{figure*}
    \centering
    \includegraphics[width=0.45\linewidth]{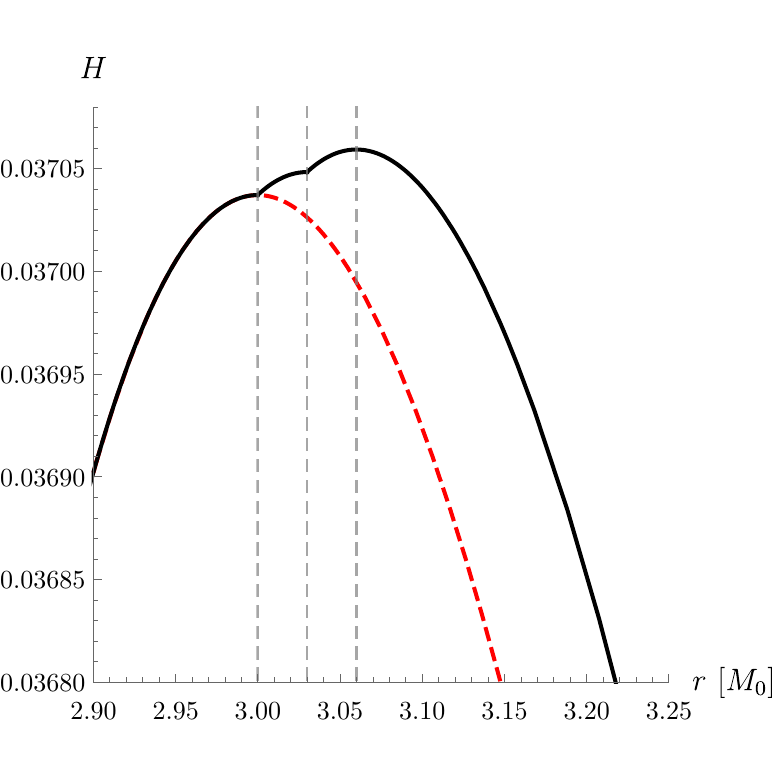}
    \hspace{0.5cm}
    \includegraphics[width=0.45\linewidth]{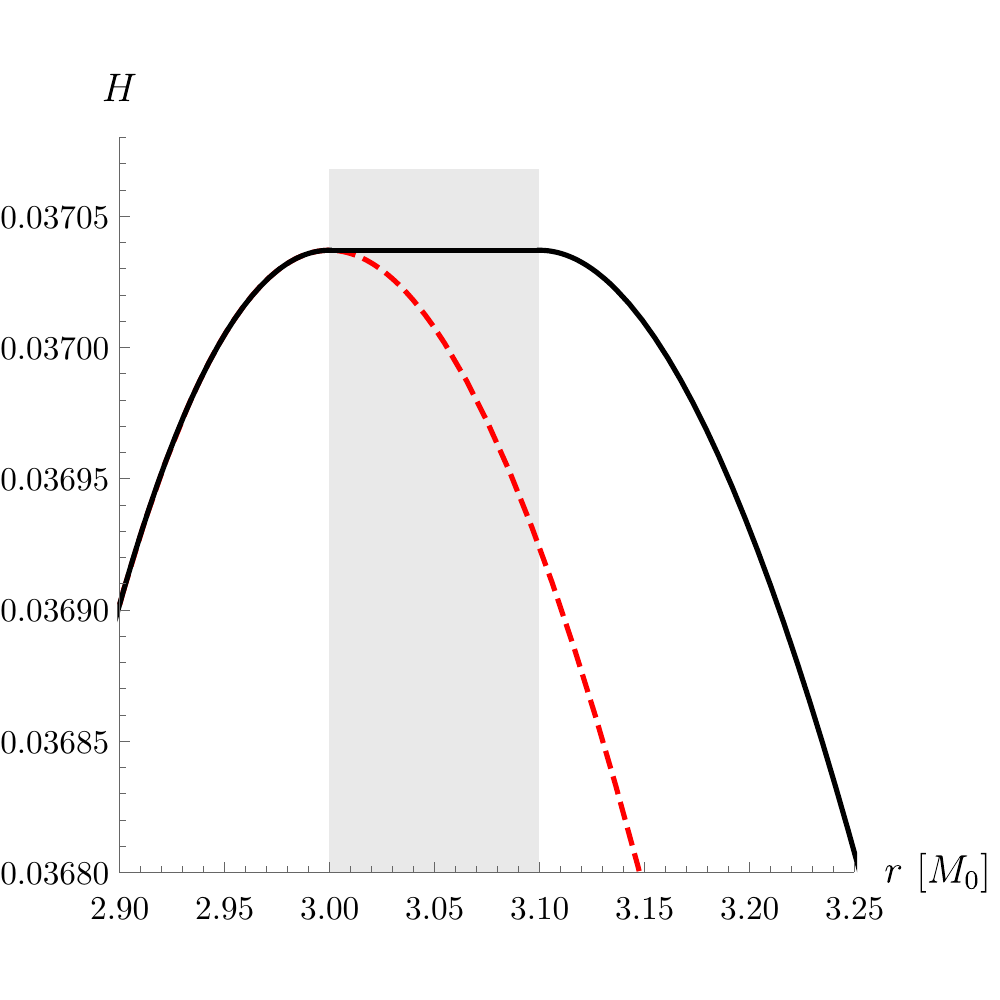}
    \caption{Shown with a black solid line is the effective potential in the case of a discrete hierarchy of light-rings (left panel, where $M_0=1$, $\delta m_1=\delta m_2=0.01\,M_0$ and the light rings are at $r_0=3\,M_0$, $r_1=3\left(M_0+\delta m_1\right)$ and $r_2=3\left(M_0+\delta m_1+\delta m_2\right)$) or a continuous sequence of light-rings (right panel,  where $M_0=1$, $R_{\rm in}=3.0\,M_0$ and $R_{\rm out}=3.1\,M_0$). The location of the light rings is shown with vertical dashed lines in the left panel and with a shaded area in the right panel. In both cases, the red dashed line shows the effective potential for a Schwarzschild black hole.}
    \label{fig:Eff_pot}
\end{figure*}

As mentioned above, $\xi^2_n$ are constants and can be fixed by simply redefining the time coordinate in each spacetime region. While this is in principle possible, it would complicate the subsequent analysis and hence we determine these constants by requiring the continuity of the time coordinate and the induced metric on the shells.
\begin{equation}\label{Eq:xi}
    \xi^2_n = \left(\frac{1-{2M_{n-1}}/{R_{ n}}}{1-{2M_{n}}/{R_{n}}}\right)\xi^2_{n-1}\,,
\end{equation} 
where $R_n$ is the radius of the $n$-th shell. Given the metric~\eqref{eq:metric_n_shells}, the corresponding effective potential for massless particles is 
\begin{equation}
    H_n(r)=\frac{1}{r^2}\xi^2_n\left( 1-\frac{2M_n}{r} \right)\,,
\end{equation}
whose extremal points marking the position of the unstable photon circular orbits  can be easily computed after imposing $\partial_r H_n(r)=0$ and are located at the $n$ different shells with coordinate radii 
\begin{equation}\label{eq:M_n(r)}
    r_n=3M_n\,.
\end{equation} 
Note that the location of the $n$-th shell coincides with the location of the light-ring of the $n-1$ spacetime region, i.e., $R_n=r_{n-1}$.
\begin{comment}
    
and this allows us to determine the conditions under which the hierarchy of light-rings actually forms a new black hole. In practice, this would happen for $M_n/R_n \geq 1/2$, that is, for 
%
\begin{equation}
    \frac{M_n}{R_n}=\frac{M_n}{3M_{n-1}}=\frac{M_{n-1}+\delta m_n}{3M_{n-1}} \geq \frac{1}{2}\,.
\end{equation}
%
Stated differently, no new black hole is formed as long as the additional mass outside the initial black hole is 
%
\begin{equation}
    \frac{\delta m_n}{M_0} \leq \frac{M_{n-1}}{2M_0} =\frac{1}{2}
    \left(1+ \frac{\delta M_{n-1}}{M_0}\right)\,,
\end{equation}
%
which is easily true since we are exploring a scenario with $\delta m_n/M_0 \ll 1$.
\end{comment}

Let us now restrict our attention to the simplest case of a single self-gravitating shell of photons on unstable circular orbits. In this case, $N=1$ and 
\begin{equation}
        \xi^2_1 = \frac{R_1-{2M_{0}}}{R_1-{2M_{1}}}\,,
    \qquad
    r_1 = 3 M_1 = 3\left(M_0 + \delta M_1\right)\,,
\end{equation}
from which 
\begin{equation}
    H_1(r)=\frac{1}{r^2}\left(\frac{R_1-{2M_{0}}}{R_1-{2M_{1}}}\right)\left( 1-\frac{2M_1}{r} \right)\,.
\end{equation}

Clearly, it is now possible to populate also the new light-ring with a second shell of carefully chosen photons and the resulting spacetime would correspond to the construction discussed above but in the case in which, $N=2$. The corresponding effective potential is shown in the left panel of Fig.~\ref{fig:Eff_pot}, where we report with a solid black line the effective potential for a black hole with $M_0=1$ and the two shells with mass $\delta m_1 = \delta m_2 = 0.01$. Note the presence of a first light-ring at $r_0=3 M_0$ and of a second one at $r_1=3 (M_0+\delta m_1)$ and a third one at $r_2=3 (M_0+\delta m_1+\delta m_2)$, where the effective potential, which is everywhere continuous but with discontinuous first derivatives at the light-rings, has extrema. In this case, only two light-rings are populated and have a nonzero mass, while the most external one has an unstable circular orbit but a zero mass. 
Clearly, as long as one is able to populate these light-rings with fine-tuned photons, the construction illustrated here can be extended simply for any integer $N$, thus creating a hierarchy of $N+1$ light-rings. 

Is this configuration stable? In principle, to address this question we would need to perform a linear perturbation analysis taking Eq.~\eqref{eq:metric_n_shells} as background spacetime and assess the properties of the imaginary eigenfrequencies (see, e.g.,~\cite{Nagar:2005ea}). In practice, we take a shortcut and note that it is not difficult to realise that such a spacetime is dynamically unstable. Consider, for instance, the addition of a single photon on any of the shells of the hierarchy. This will be sufficient to change the mass of the corresponding shell and hence place on unstable photon orbits all the photons on the light-rings that are at larger radii than the one on which the photon is deposited. Similarly, if a single photon is absorbed by the black hole, this would destabilise all of the light-rings, dispersing or absorbing all the photons that have been accumulated. Clearly, while we believe this conclusion is correct, a proper perturbative analysis is needed to confirm it.

%------------------------------------------------------
%\section{Continuous sequence of light-rings}
%\label{sec:self-grav-continuum}
%------------------------------------------------------

\smallskip
\noindent\textit{Continuous sequence of light-rings.~} As mentioned in the introduction, the ``Gedankenexperiment" can be carried out also in a different way. More specifically, rather than accumulating photons on the first light-ring till they produce a discrete contribution to the spacetime and generate a new light-ring, we can imagine a situation in which the advanced civilization concentrates on adding photons in the right amount, both in terms on numbers and impact parameters, so that the spacetime properties are varied smoothly. This would correspond to a situation in which both the mass of each individual shell and the distance among shells goes to zero while the total mass of the shells remains finite. This configuration would then describe a continuous energy distribution within a finite region of light-rings, i.e., a ``light-region'' with radius $r\in \left[R_{\rm in}, R_{\rm out}\right]$, with $R_{\rm in}=3\,M_0$.

Hence, we start by replacing the discrete nature of the mass in the light-ring %given by Eq.~\eqref{eq:M} 
with the continuous function $M(r) = M_0 + \delta M(r)$, and 
where $\delta M(r)$ denotes the full energy of the perturbation within a radius $r$, so that $\delta M(R_{\rm in})=0$. Furthermore, as $M(r)$ denotes the total energy within a radius $r$. Also in this case we can use the coefficients $\xi_n$ and calculate their values in the continuous limit. More specifically, we can rewrite Eq.~\eqref{Eq:xi} as 
\begin{eqnarray}
\hspace*{-.4cm}
    \label{eq:xi2_n_O2}
    \xi^2_n =&\displaystyle \left(\frac{R_n-{2M_{n-1}}}{R_n-{2M_{n}}}\right)\xi^2_{n-1} =  
    \prod_{i=1}^n\left(1+\frac{2\delta m_i}{R_i-{2M_{i}}}\right).
\end{eqnarray}
The continuous limit of this product is known and it is given by a Volterra-type integral\footnote{We are indebted to the anonymous referee for pointing out the possibility of employing an integral representation.}. The generic expression is (see e.g.~\cite{Slavík2007})
\begin{equation}
    \lim_{\Delta X\rightarrow0}\prod_{i=0}^N\left(1+f(x_i)\Delta x\right)=\exp{\int_{x_0}^{x_N}f(x){\rm d}x}\,,
\end{equation}
that, in our case, yields
\begin{equation}
    \xi^2(r) =\exp{\displaystyle\int_{3M_0}^r\frac{2 M'(\bar{r})}{\bar{r}-2M(\bar{r})}{\rm d}\bar{r}}\,.
\end{equation}
We can easily compute the integral using the countinuous version of Eq.~\eqref{eq:M_n(r)}, namely $M(r)=r/3$, obtaining
\begin{equation}
        \xi^2(r) =\exp{\displaystyle\int_{3M_0}^r\frac{2}{\bar{r}}{\rm d}\bar{r}}=\left(\frac{r}{3M_0}\right)^2\,.
\end{equation}
Therefore, the effective potential is given by
\begin{equation}
    H(r) =
    \begin{cases}
      \displaystyle\frac{1}{r^2}\left(1-\frac{2M_0}{r}\right) &  r < R_{\text{in}} \\
      \\
      \displaystyle\frac{1}{27 M_0^2} &  R_{\text{in}} \leq r \leq R_{\text{out}} \\
      \\
     \displaystyle \frac{R_{\rm out}^2}{9M_0^2 r^2}\left(1-\frac{2M\left(R_{\rm out}\right)}{r}\right) & r > R_{\text{out}} \\
    \end{cases}
\end{equation}

In summary, we have shown that it is possible to create a self-gravitating ``light-region" that extends from the position of the original light-ring $R_{\rm in}$ up to $R_{\rm out}$.
The corresponding potential is shown with a black solid line in the right panel of Fig.~\ref{fig:Eff_pot} and it is then easy to appreciate how we have essentially created a smooth region of light-rings where the potential is flat in the region $R_{\rm in} \leq r \leq R_{\rm out}$, hence leading to an infinite sequence of \textit{neutral} circular photon orbits in such a region, i.e., with zero first and second derivative of the effective potential.

We can now ask the same question: is this configuration stable? Obviously, a photon on a circular orbit well inside the light-region, i.e., at  $R_{\rm in} \lesssim r \lesssim R_{\rm out}$, and that is perturbed, would still encounter in its neighbourhood other circular orbits and hence remain in the newly and perturbed location. However, it is also easy to conclude that the whole light-region is dynamically unstable after a small change in the mass of the black hole. Consider in fact that the mass of the spacetime is changed by a small but finite amount $\delta M_0$ without changing the mass in the light-rings, namely, consider the new mass profile $M(r)=M_0+\delta M_0+\delta M(r)$ with both $M_0$ and $\delta M_0$ constants. After straightforward computations, the new effective potential, in the range  $r\in \left[R_{\rm in}, R_{\rm out}\right]$, will be

\begin{align}
    H(r)&=\frac{1}{27M_0^2}-\frac{2  (6 M_0-r)}{27 M_0^3 r}\delta M_0+\mathcal{O}(\delta M_0)^2\,.
 \end{align}
So the radial derivative no longer vanishes but
\begin{equation}
    \partial_r H(r)= \frac{4}{9M_0^2 r^2}\delta M_0+\mathcal{O}(\delta M_0)^2>0\,. 
\end{equation}
Stated differently, a small perturbation in the central black hole would completely destabilise the circular orbits in the light-region, dispersing all the photons carefully accumulated there.

Finally, when looking at the left panel of  Fig.~\ref{fig:Eff_pot}, where each shell increases the value of the maximum of the potential, one might expect that in the continuous limit the slope of the potential would be nonzero in the shaded region of the right panel of Fig.~\ref{fig:Eff_pot}. However, this is not the case, as the discrete slope $[H(r_n)-H(r_{n-1})]/(r_n-r_{n-1})$ vanishes in the continuous limit. Indeed, a straightforward computation shows that $H(r_n)-H(r_{n-1})\propto \left(\delta m_n\right)^2$ while $r_n - r_{n-1} \propto \delta m_n$, so that their ratio is zero for $\delta m_n \to 0$.

%------------------------------------------------------
%\section{Conclusion}
%\label{sec:conclusion}
%------------------------------------------------------

\smallskip
\noindent\textit{Conclusion.~} Light-rings represent a powerful tool to probe the spacetime of black holes and test different theories of gravity, both in terms of electromagnetic and gravitational radiation. So far, however, the self-gravity of light-rings has been neglected on the assumption that photons entering the light-ring are compensated by photons leaving, so that the corresponding mass is not sufficient to modify the spacetime appreciably. We have here explored the scenario in which this assumption is not true the light-ring of a Schwarzschild black hole is filled with carefully injected photons making it ``self-gravitate''.

%\noindent\textit{Conclusion.~} Light-rings represent one of the most interesting aspects of the spacetime around compact objects in general and of black holes in particular. Being characterised by circular but unstable orbits, they represent a powerful tool to probe the spacetime near these objects and test different theories of gravity, both in terms of electromagnetic and gravitational radiation. So far, however, the self-gravity of light-rings has been neglected and this is because one expects that, on average, the number of photons entering the light-ring equals that of photons leaving the light-ring, and that such a number is not sufficient to modify the spacetime appreciably. We have here explored what would happen when this assumption is not true and for this we considered ways in which the light-ring in a Schwarzschild spacetime can be filled with carefully injected photons so as to make the light-ring self-gravitate. 

In particular, we have considered two different routes for assigning a nonzero mass to the light-ring and that are either based on a discrete concentration of photons on a specific radial location or on suitable distribution of photons in a given region. Notwithstanding a number of differences between the two scenarios, we have found that, in both cases, the inclusion of the energy contribution from the accumulated photons leads to the generation of new light-rings. Such new light-rings can either appear at well-defined but discrete locations, or be fused with zero separation in a well-defined region. Interestingly, and somewhat disappointingly, all light-ring configurations considered here are unstable and a small perturbation, either via the inclusion on an additional photon onto the light-ring or via the absorption of a photon by the black hole, leads to a catastrophic destruction of the light-ring structures and to the dispersion of the photons to spatial infinity or inside the black hole. Hence, the answer to the question in the title is: ``yes, light-rings can self-gravitate, but they are dynamically unstable".

This work could be extended in several ways. First, by performing a perturbative analysis that would also provide the timescale of the instability. Second, by using these analytical spacetimes to study the quasi-normal modes or the emergence of echos. Third, by considering Kerr spacetimes and the corresponding light-rings. Furthermore, much of the work started can be extended to horizonless ultracompact objects as long as they possess an unstable light-ring a hence also stable light-rings~\cite{Mazur2001, Cunha:2017qtt, Guo:2020qwk, Jampolski2024, DiFilippo:2024mnc}.
Finally, while the scenarios explored in this paper are so idealised to require the intervention of an advanced civilisation, future investigations might consider more realistic sources of photons. Even without fine-tuning the impact parameter, photons with an impact parameter close enough to populate the light-ring will orbit the black hole several times, spending on average more time in this region. This would naturally lead to an over-density at the light-ring similar to the one considered in our work.

\bigskip
\acknowledgements

It is a pleasure to thank R. Carballo-Rubio, P. Cunha, C. Ecker, C. Herdeiro, P. Kocherlakota, R. Kumar Walia, D. Jampolski, S. Liberati, H. Olivares, and P. Schupp for interesting discussions and suggestions. FDF acknowledges financial support from the PRIMUS/23/SCI/005 and UNCE24/SCI/016 grants by Charles University, and the GAR 23-07457S grant from the Czech Science Foundation. LR acknowledges financial support from the State of Hesse within the Research Cluster ELEMENTS (Project ID 500/10.006), from the ERC Advanced Grant ``JETSET: Launching, propagation and emission of relativistic jets from binary mergers and across mass scales'' (Grant No. 884631), and from the Walter Greiner Gesellschaft zur F\"orderung der physikalischen Grundlagenforschung e.V. through the Carl W. Fueck Laureatus Chair.
\bibliography{ref}
\end{document}